\documentclass[12pt]{article}
\usepackage{geometry}
\usepackage{here}
\usepackage{graphicx}
\usepackage{amsmath,amssymb}
\usepackage{cite}
\usepackage{bm}
\newcommand{\lw}[1]{\smash{\lower 1.5ex\hbox{#1}}}
\newcommand{\ri}[1]{\smash{\raise 1.5ex\hbox{#1}}}
\newcommand{\riw}[1]{\smash{\raise 3.0ex\hbox{#1}}}

\newcommand{\mapright}[1]{\smash{\mathop{\hbox to 3.0cm{\rightarrowfill}}\limits^{\displaystyle #1}}}

\begin{document}

\title{Analysis of OPAL Bose-Einstein Correlation at $Z^0$-pole by the second conventional formula}

\author{Takuya Mizoguchi$^{1}$, Seiji Matsumoto$^{2}$, and Minoru Biyajima$^{3}$\\
{\small $^{1}$National Institute of Technology, Toba College, Toba 517-8501, Japan}\\
{\small $^{2}$School of General Education, Shinshu University, Matsumoto 390-8621, Japan}\\
{\small $^{3}$Department of Physics, Shinshu University, Matsumoto 390-8621, Japan}}

\maketitle

\begin{abstract}
We can obtain a second conventional formula (${\rm CF_{II}}$) with two components by extending the conventional formula ${\rm CF_{I}}$ with one component for Bose-Einstein Correlation (BEC). We used ${\rm CF_{II}}$ to analyze the BEC at $Z^0$-pole as part of the OPAL collaboration. $R_1({\rm G})=0.91\pm 0.03$ fm ($\lambda_1=0.61\pm 0.03$) and $R_2({\rm G})=2.57\pm 0.44$ fm ($\lambda_1=0.35\pm 0.09$) are the estimated interaction region, where $\lambda_1$ and $\lambda_2$ are the degrees of coherence, and G is the Gaussian distribution, respectively. Long range correlation (LRC) is also studied.
\end{abstract}

\section{\label{sec1}Introduction}
In the analysis of Bose-Einstein Correlation (BEC)~\cite{HanburyBrown:1956,Acton:1991xb,Abreu:1992gj}, the following conventional formula is commonly used,
\begin{eqnarray}
{\rm CF_{I}\cdot LRC} = \left[1.0 + \lambda E_{\rm BE}(R,\,Q)\right]\cdot {\rm LRC},
\label{eq1}
\end{eqnarray}
where $E_{\rm BE}$ is the exchange function between two identical charged pions, which is expressed by 
\begin{eqnarray}
E_{\rm BE} =\ \left\{
\begin{array}{l}
\exp(-(RQ)^2) \mbox{ (Gaussian distribution (G))},\medskip\\
\mbox{or }\exp(-RQ) \mbox{ (Exponential function (E))},
\end{array}
\right.
\label{eq2}
\end{eqnarray}
$\lambda$ is the degree of coherence. $Q=\sqrt{(p_1-p_2)^2}=\sqrt{(\bm p_1-\bm p_2)^2-(E_1-E_2)^2}$ is the four momentum transfer squared between two identical pions momenta. LRC is the long range correlation, which is typically expressed by
\begin{eqnarray}
{\rm LRC} =\ \left\{
\begin{array}{l}
1.0 + \delta Q,\medskip\\
\mbox{or }1.0 + \delta Q + \varepsilon Q^2.
\end{array}
\right.
\label{eq3}
\end{eqnarray}

\begin{figure}[htbp]
  \centering
  \includegraphics[width=0.48\columnwidth]{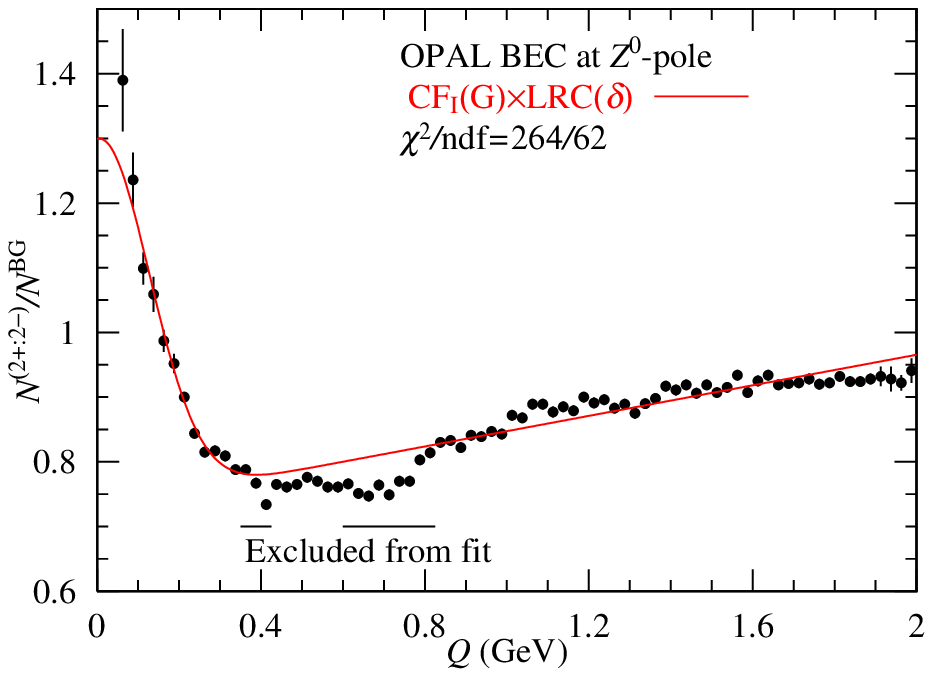}
  \includegraphics[width=0.48\columnwidth]{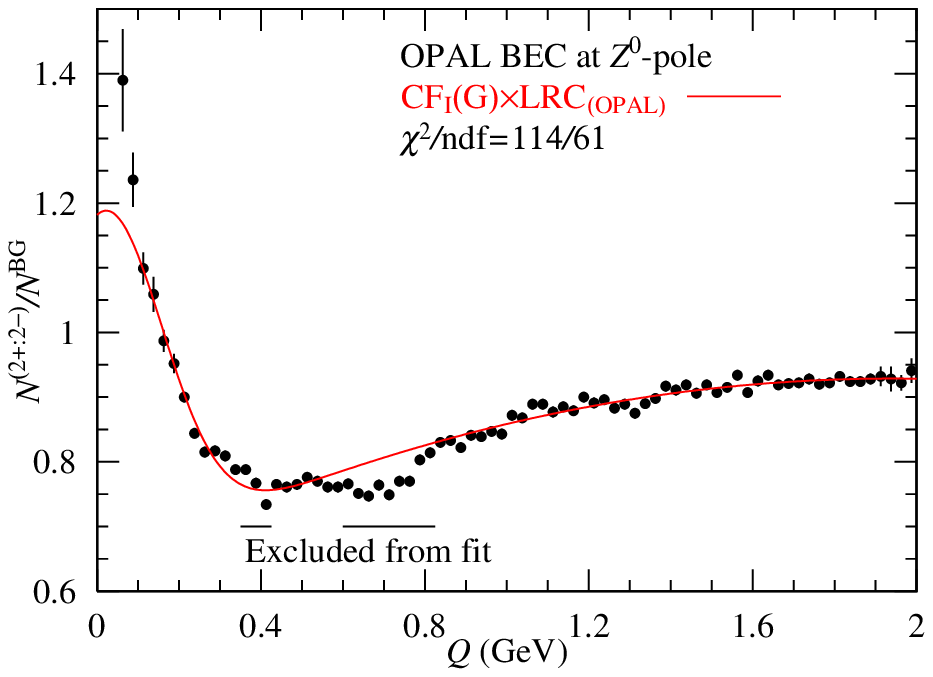}
  \caption{\label{fig1}Analysis of the OPAL BEC at the $Z^0$-pole by Eq.~(\ref{eq1}) with Eqs.~(\ref{eq2}) and (\ref{eq3}).}
\end{figure}

The OPAL collaboration used ${\rm CF_{I}(G)\times LRC}$~\cite{Acton:1991xb} to analyze their data. The results are shown in Table~\ref{tab1} and Fig.~\ref{fig1}.

\begin{table}[H]
\centering
\caption{\label{tab1}Analysis of OPAL data using Eqs.~(\ref{eq1})--(\ref{eq3}). (G) denotes the Gaussian distribution. $p$(\%) denotes the p-value in the statistics.}
\vspace{1mm}
\begin{tabular}{c|ccccc}
\hline
LRC 
& $R$ (fm) 
& $\lambda$ 
& $C$ 
& $\delta$ (GeV$^{-1}$) 
& $\varepsilon$ (GeV$^{-2}$)\\
\hline

\lw{($\delta$)}
& $1.115\pm 0.026$ (G)
& $0.782\pm 0.039$
& $0.729\pm 0.004$
& $0.162\pm 0.004$
& ---\\
& \multicolumn{5}{l}{$\chi^2/{\rm ndf}=263.8/62$ ($p\,(\%)=0.0$)}\\
\hline

\lw{($\delta$, $\varepsilon$)}
& $ 0.949\pm 0.024$ (G)
& $ 0.876\pm 0.037$
& $ 0.630\pm 0.010$
& $ 0.489\pm 0.034$
& $-0.126\pm 0.012$\\
& \multicolumn{5}{l}{$\chi^2/{\rm ndf}=113.9/61$ ($p\,(\%)=0.0046$)}\\
\hline
\end{tabular}
\end{table}

\begin{figure}[H]
  \centering
  \vspace{0mm}
  \includegraphics[width=0.32\columnwidth]{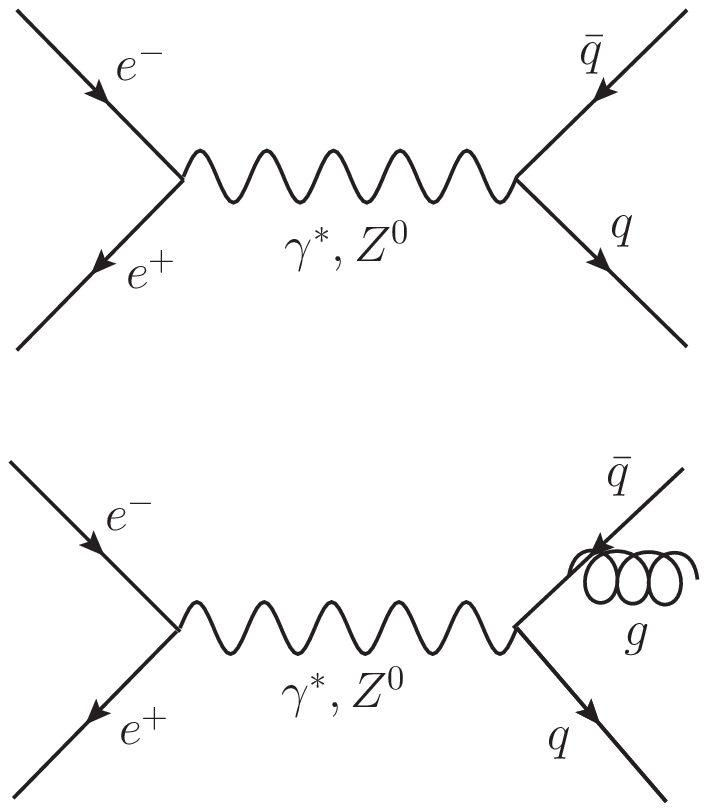}
  \includegraphics[width=0.23\columnwidth]{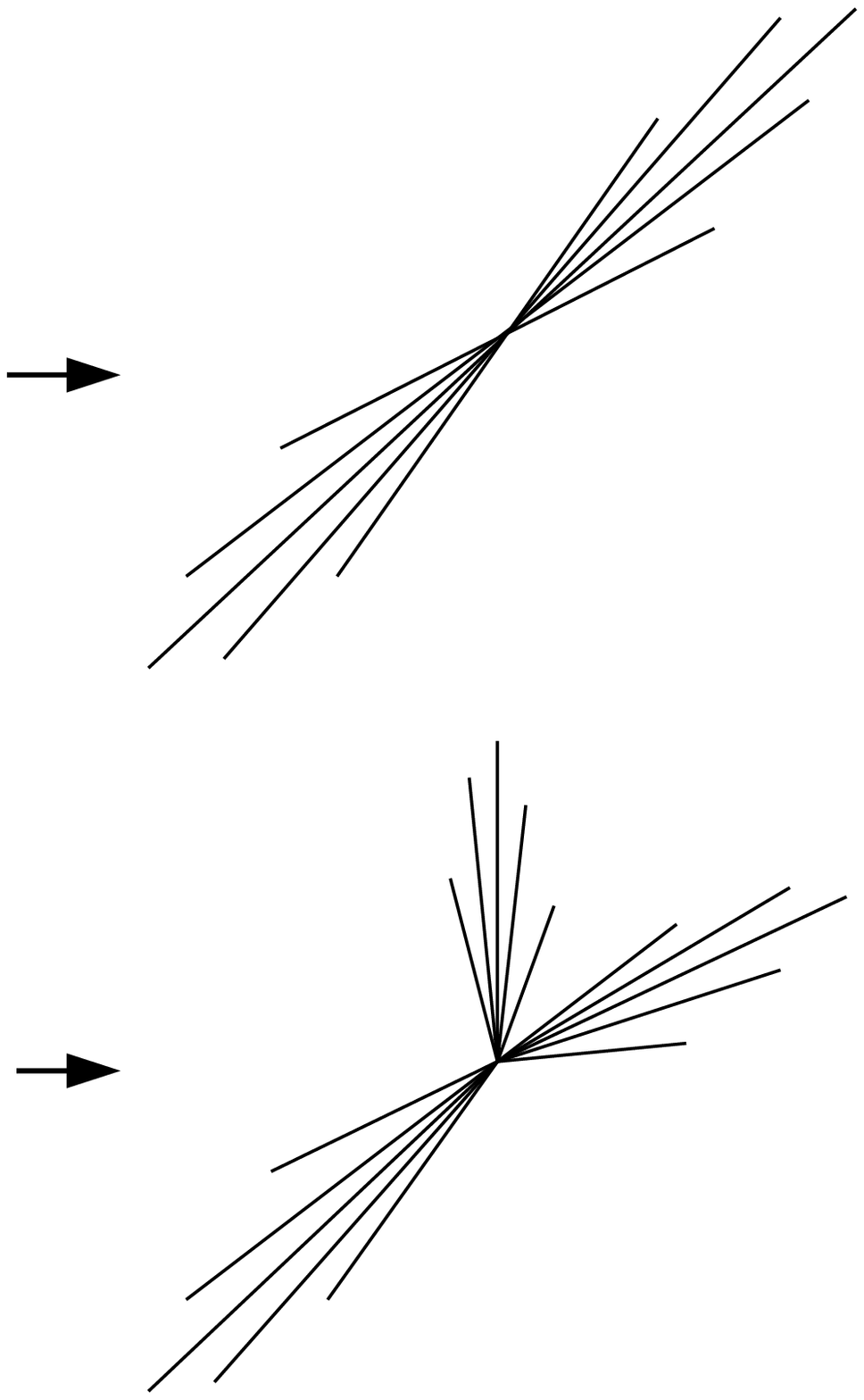}
  \vspace{0mm}
  \caption{\label{fig2}Jet-production in $e^+e^-$ annihilation.}
  \vspace{0mm}
\end{figure}

In Refs.~\cite{Aad:2015sja,Khachatryan:2011hi,Biyajima:2018abe,Mizoguchi:2019cra,Biyajima:2019wcb}, we analyzed the BEC at 0.9 TeV and 7 TeV by CMS collaboration using the CF$_{\rm I}$ formula. In those analyses, $\chi^2/$ndf's are very large. To improve them, we use the CF$_{\rm II}$ formula, which corresponds to the nondiffractive and diffractive process in $pp$ collisions with two components,
\begin{eqnarray}
{\rm CF_{II}\cdot LRC} = \left[(1.0 + \lambda_1 E_{\rm BE_1}(R_1,\,Q) + \lambda_2 E_{\rm BE_2}(R_2,\,Q)\right]\cdot {\rm LRC_{(OPAL)}}.
\label{eq4}
\end{eqnarray}
Indeed, $\chi^2/$ndf's became smaller values. Thus, we expect that Eq.~(\ref{eq4}) also works for the BEC at $Z^0$-pole because there are two types of events in $e^+e^-$ annihilation. See Fig.~\ref{fig2}~\cite{Foster:1990,Mizoguchi:2020,Giovannini:1996,Duchesnean:2002}.

In section~\ref{sec2}, we analyze the BEC at $Z^0$-pole by ${\rm CF_{II}\times LRC_{(OPAL)}}$, because there are the 2-jet and 3-jet events in $e^+e^-$ collisions at $Z^0$-pole. In section~\ref{sec3}, we investigate the LRC$_{\rm (OPAL)}$ from a slightly different angle, considering the behavior in the large and small $Q$ regions. In section \ref{sec4}, the profile of LRC's and the exchange functions in the Euclidean space is studied. In the final section, we present the concluding remarks.

\section{\label{sec2}Analysis of BEC using CF$_{\rm II}\times$LRC$_{\rm (OPAL)}$}
First, we focus on the results in Table~\ref{tab1}. Because they are considered the first approximation, we adopt the following constraints for $R_1$, $\lambda_1$, $c$, $\delta$, and $\varepsilon$:
\begin{eqnarray}
q-\sigma_q\le q\le q+\sigma_q\ \Rightarrow\ \left\{
\begin{array}{l}
 0.925 \le R_1 \le 0.973\ ({\rm fm}),\medskip\\
 0.840 \le \lambda_1 \le 0.914,\medskip\\
 0.620 \le C \le 0.640,\medskip\\
 0.455 \le \delta \le 0.523\ ({\rm GeV}^{-1}),\medskip\\
-0.138 \le \varepsilon \le -0.114\ ({\rm GeV}^{-2}).
\end{array}
\right.
\label{eq5}
\end{eqnarray}
Moreover, the following three constraints are introduced (Fig.~\ref{fig2}).
\begin{eqnarray}
\lambda_2 < \lambda_1,\ \lambda_1 + \lambda_2\le 1.0,\ {\rm and}\ R_1 < R_2.
\label{eq6}
\end{eqnarray}
In Fig.~\ref{fig2}, $R_2$ and $\lambda_2$ denote the interaction region and the second degree of coherent of 3-jet, respectively.

\begin{figure}[H]
  \centering
  \vspace{0mm}
  \includegraphics[width=0.6\columnwidth]{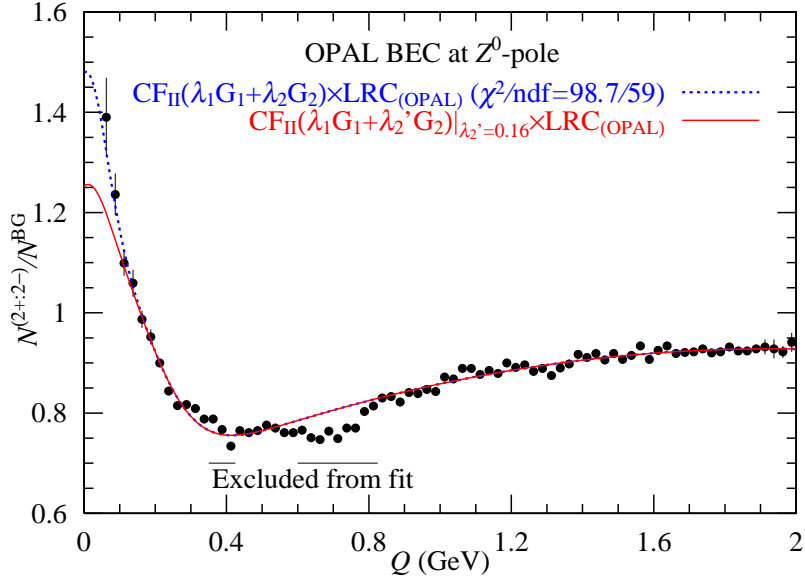}
  \vspace{0mm}
  \caption{\label{fig3}Analysis of the OPAL BEC at the $Z^0$-pole by Eq.~(\ref{eq4}) with no-constraint for $\lambda_1$ and $\lambda_2$. ${\rm LRC_{(OPAL)}}=1.0 + \delta Q + \varepsilon Q^2$. See Table~\ref{tab3}.}
  \vspace{0mm}
\end{figure}

${\rm CF_{II}}(\lambda_1{\rm G_1}+\lambda_2{\rm G_2})$ and ${\rm CF_{II}}(\lambda_1{\rm G_1}+\lambda_2{\rm E_2})$ are two plausible geometrical combinations. Among them, the ${\rm CF_{II}}(\lambda_1{\rm G_1}+\lambda_2{\rm G_2})$ has the minimum $\chi^2$/ndf. Thus, our result by ${\rm CF_{II}}(\lambda_1{\rm G_1}+\lambda_2{\rm G_2})\cdot {\rm LRC_{(OPAL)}}$ is shown in Tables~\ref{tab2} and \ref{tab3}, as well as Fig.~\ref{fig3}.

\begin{table}[H]
\centering
\caption{\label{tab2}Estimated parameters using Eq.~(\ref{eq4}) with the geometrical combination ($\lambda_1{\rm G_1}+\lambda_2{\rm G_2}$)and the constraint $\lambda_1 + \lambda_2\le 1.0$.} 
\vspace{2mm}
\renewcommand{\arraystretch}{1.0}
\begin{tabular}{cccc}
\hline
  $R_1$ (fm) (G)
& $R_2$ (fm) (G)
& $\lambda_1$
& $\lambda_2$\\
  $ 0.925\pm 0.048$
& $ 1.926\pm 0.277$
& $ 0.840\pm 0.050$
& $ 0.160\pm 0.024$\\
\hline

  $C$
& $\delta$ (GeV$^{-1}$) 
& $\varepsilon$ (GeV$^{-2}$) 
& $\chi^2$/ndf ($p$(\%))\\
  $ 0.624\pm 0.004$
& $ 0.509\pm 0.017$
& $-0.133\pm 0.007$
& 104.9/59 (0.022)\\
\hline
\end{tabular}
\end{table}

\begin{table}[H]
\centering
\caption{\label{tab3}Estimated parameters using Eq.~(\ref{eq4}) with the geometrical combination ($\lambda_1{\rm G_1}+\lambda_2{\rm G_2}$). Although $\lambda_1 + \lambda_2$ is not bounded, the assumption $\lambda_1 + \lambda_2'=1.0$ is used.  $\Delta$ denotes the difference between the dot-dot line and solid line in Fig. \ref{fig3}. The OPAL collaboration observed that $\Delta = 0.15$~\cite{Acton:1991xb}.} 
\vspace{2mm}
\renewcommand{\arraystretch}{1.0}
\begin{tabular}{cccc}
\hline
  $R_1$ (fm) (G)
& $R_2$ (fm) (G)
& $\lambda_1$
& $\lambda_2$\\
  $ 0.931\pm 0.009$
& $ 2.506\pm 0.339$
& $ 0.841\pm 0.010$
& $ 0.521\pm 0.159$\\
\hline

  $C$
& $\delta$ (GeV$^{-1}$) 
& $\varepsilon$ (GeV$^{-2}$) 
& $\chi^2$/ndf ($p$(\%))\\
  $ 0.626\pm 0.006$
& $ 0.501\pm 0.021$
& $-0.130\pm 0.008$
& 98.7/59 (0.091)\\
\hline
\end{tabular}

\begin{tabular}{c|ccc}
\hline
Note
& $R_2$ (fm)
& $\lambda_2'$
& $\lambda_2''$\\

\lw{``$\lambda_1+\lambda_2'=1.0$''}
& $ 2.51\pm 0.34$
& $ 0.16\pm 0.01$
& $ 0.36\pm 0.16$\\

\lw{is assumed.}
&& 2nd-BEC from 
& Contribution from $\eta'$ and $\eta$\\
&& 3-jet.
&  decay chain, $\Delta \cong 0.13$.\\
\hline
\end{tabular}

\end{table}

\section{\label{sec3}Different expressions of LRC$_{\rm (OPAL)}$}
%
\subsection{\label{sec3-1}Inclusion of higher order of variable $Q$ in LRC}
\begin{figure}[H]
  \centering
  \vspace{0mm}
  \includegraphics[width=0.75\columnwidth]{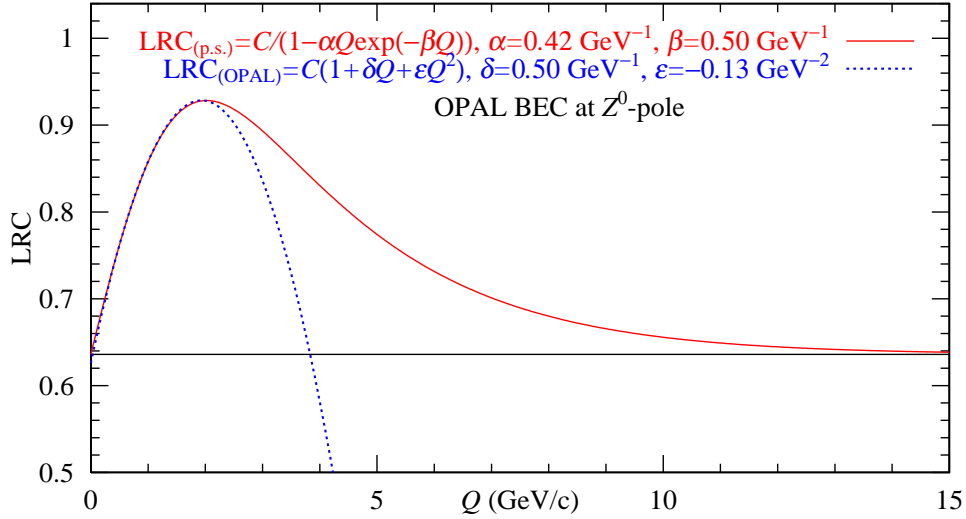}
  \vspace{0mm}
  \caption{\label{fig4}LRCs of the OPAL BEC at the $Z^0$-pole by Eqs.~(\ref{eq4}) and (\ref{eq8}). Numerical values are presented in Tables~\ref{tab3} and \ref{tab4}.}
  \vspace{0mm}
\end{figure}

Fig.~\ref{fig4} shows the behavior of LRC$_{\rm (OPAL)}$. We observe that LRC$_{\rm (OPAL)}$ becomes $C=0.63$ at $Q\cong 4$ GeV. Indeed that behavior is a parabolic curve.

Here, we introduce a function $X = \alpha Q e^{-\beta Q}=\alpha Q(1-\beta Q+\beta^2 Q^2/2!-\cdots)$ and it is a power series because $\alpha$ and $\beta$ are two parameters, as shown below.
\begin{eqnarray}
{\rm LRC_{(p.s.)}} &\!\!\!=&\!\!\! C\left[\sum_{k=0}^{\infty} \left(\alpha Q e^{-\beta Q}\right)^k \right],\nonumber\\
&\!\!\!=&\!\!\! \frac C{1-\alpha Q e^{-\beta Q}}.
\label{eq7}
\end{eqnarray}

In Eq.~(\ref{eq7}), we can avoid the following behavior in ${\rm LRC_{(OPAL)}}$ (in the region $Q>4.0$ GeV),
$$
{\rm LRC_{(OPAL)}} \mapright{Q\to {\rm large}}\ \mbox{negative values.}
$$

Fig.~\ref{fig4} compares two LRCs, ${\rm LRC_{(OPAL)}}$ and ${\rm LRC_{(p.s.)}}$. The Fig. coincidence between ${\rm LRC_{(OPAL)}}$ and ${\rm LRC_{(p.s.)}}$ in the region $Q\le 2.0$ GeV seems to be excellent. Moreover, ${\rm LRC_{(p.s.)}}$ gradually becomes the constant values $C$ at large $Q$.

\begin{figure}[H]
  \centering
  \vspace{0mm}
  \includegraphics[width=0.6\columnwidth]{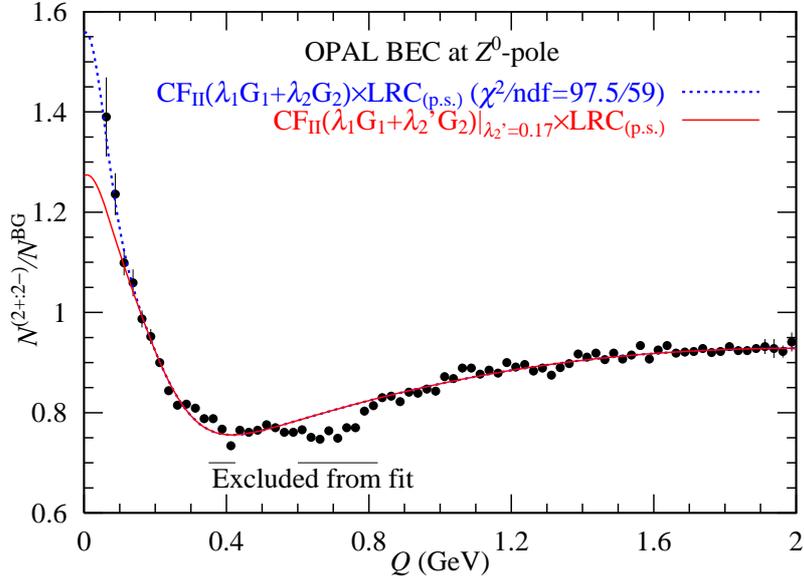}
  \vspace{0mm}
  \caption{\label{fig5}Analysis of the OPAL BEC at the $Z^0$-pole by Eq.~(\ref{eq8}). ${\rm LRC_{(p.s.)}} =\sum_{k=0}^{\infty} \left(\alpha Q e^{-\beta Q}\right)^k$. See Table~\ref{tab4}.}
  \vspace{0mm}
\end{figure}

This time the following  formula is used in our analysis.
\begin{eqnarray}
{\rm CF_{II}\times LRC_{(p.s.)}} = (1.0+\lambda_1 E_{\rm BE_1}(R_1,\,Q)+\lambda_2 E_{\rm BE_2}(R_2,\,Q))\cdot {\rm LRC_{(p.s.)}}
\label{eq8}
\end{eqnarray}
Our result from Eq.~(\ref{eq8}) is presented in Fig.~\ref{fig5} and Table~\ref{tab4}, and it is almost the same as those shown in Table~\ref{tab3}.

\begin{table}[H]
\centering
\caption{\label{tab4}Estimated parameters using Eq.~(\ref{eq8}) with the geometrical combination ($\lambda_1\times {\rm G_1}+\lambda_2\times {\rm G_2}$). Although $\lambda_1 + \lambda_2$ is not bounded, the assumption $\lambda_1 + \lambda_2'=1.0$ is used. $\Delta$ denotes the difference between the dot-dot line and solid line in Fig. \ref{fig5}. The OPAL collaboration observed that $\Delta = 0.15$~\cite{Acton:1991xb}.} 
\vspace{2mm}
\renewcommand{\arraystretch}{1.0}
\begin{tabular}{cccc}
\hline
  $R_1$ (fm) (G)
& $R_2$ (fm) (G)
& $\lambda_1$
& $\lambda_2$\\
  $ 0.930\pm 0.010$
& $ 2.659\pm 0.361$
& $ 0.826\pm 0.008$
& $ 0.623\pm 0.205$\\
\hline

  $C$
& $\alpha$ (GeV$^{-1}$) 
& $\beta$ (GeV$^{-1}$) 
& $\chi^2$/ndf ($p$(\%))\\
  $ 0.636\pm 0.005$
& $ 0.424\pm 0.016$
& $ 0.495\pm 0.014$
& 97.5/59 (0.12)\\
\hline
\end{tabular}

\begin{tabular}{c|ccc}
\hline
Note
& $R_2$ (fm)
& $\lambda_2'$
& $\lambda_2''$\\

\lw{``$\lambda_1+\lambda_2'=1.0$''}
& $ 2.66\pm 0.36$
& $ 0.17\pm 0.01$
& $ 0.45\pm 0.21$\\

\lw{is assumed.}
&& 2nd-BEC from 
& Contribution from $\eta'$ and $\eta$\\
&& 3-jet.
&  decay chain, $\Delta \cong 0.15$.\\
\hline
\end{tabular}

\end{table}

\subsection{\label{sec3-2}Separation of $\eta'$ and $\eta$ decay chain using a modification of LRC$_{\rm (p.s.)}$}
In a previous subsection, we included the resonance effect due to $\eta'$ and $\eta$ decay chain into ${\rm CF_{II}}$ without the constraint $\lambda_1 + \lambda_2 \le 1.0$. Hereafter, we include the resonance effect in the LRC as follows.

\begin{figure}[H]
  \centering
  \vspace{0mm}
  \includegraphics[width=0.55\columnwidth]{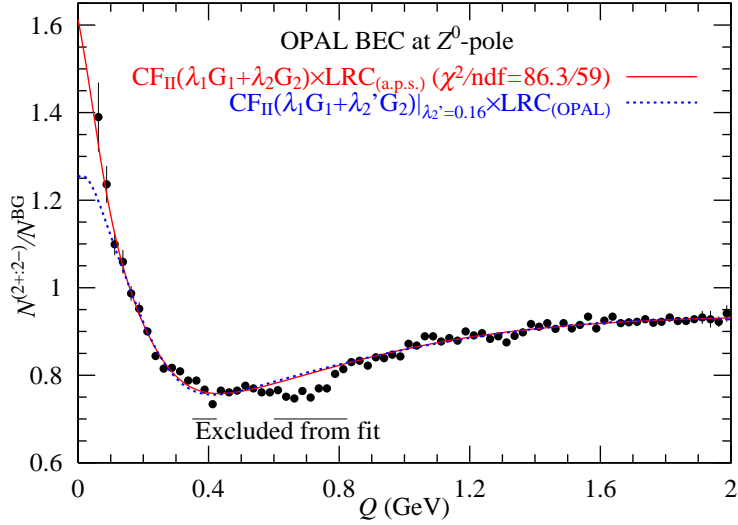}
  \vspace{0mm}
  \caption{\label{fig6}Analysis of the OPAL BEC at the $Z^0$-pole by Eq.~(\ref{eq10}).}
  \vspace{0mm}
\end{figure}

The power series in Eq.~(\ref{eq7}) is $\sum_{k=1}^{\infty} X^k$, which is an increasing function in the region $Q\le 2.0$ GeV. Here we adopt the following alternating series ${\rm LRC_{(a.p.s.)}} = \sum_{k=0}^{\infty} (-X)^k = 1.0 - X + X^2 - X^3 + X^4 - X^5 + X^6 - \cdots$, 
\begin{eqnarray}
{\rm LRC_{(a.p.s.)}} &\!\!\!=&\!\!\! C\left[\sum_{k=0}^{\infty} \left(-\alpha Q e^{-\beta Q}\right)^k \right]\nonumber\\
&\!\!\!=&\!\!\! \frac C{1+\alpha Q e^{-\beta Q}}
\label{eq9}
\end{eqnarray}
because it is the decreasing function from $Q=0.0$ GeV and increasing from $Q = 1/\beta$ GeV, implying the local minimum point. Thus, we can expect a possibility in Eq.~(\ref{eq9}) as a role of an alternative LRC. 
\begin{eqnarray}
{\rm CF_{II}\times LRC_{(a.p.s.)}} = (1.0+\lambda_1 E_{\rm BE_1}(R_1,\,Q)+\lambda_2 E_{\rm BE_2}(R_2,\,Q))\cdot {\rm LRC_{(a.p.s.)}}
\label{eq10}
\end{eqnarray}
Table~\ref{tab5} and Fig.~\ref{fig6} show our result from Eq.~(\ref{eq10}). Note that the estimated $\lambda_1 + \lambda_2 = 0.49 + 0.21 = 0.70 < 1.0$ is less than the bind value because ${\rm LRC_{(a.p.s.)}}$ (Eq. (\ref{eq9})) absorbed the resonance effect. By using estimated parameters, we can calculate the contributions of 2-jet and 3-jet as follows;
\begin{eqnarray*}
S_2 &\!\!\!=&\!\!\! \mbox{contribution of 2-jet} = \frac{\lambda_1\sqrt{\pi}}{2\times R_1\times 5.0} = 0.093,\\
S_3 &\!\!\!=&\!\!\! \mbox{contribution of 3-jet} = \frac{\lambda_2\sqrt{\pi}}{2\times R_2\times 5.0} = 0.015.
\end{eqnarray*}
The ratio of $S_3$ to all-jet is obtained as follows:
\begin{eqnarray*}
R_{\rm 3\mathchar`-all\mathchar`-jet}=\frac{S_3}{S_2+S_3} = \frac{0.015}{0.108} = 0.14 = 14\%
\end{eqnarray*}
The ratio $R_{\rm 3\mathchar`-all\mathchar`-jet}=14\%$ is compatible with the emperical value $17.4\pm 0.7\%$ in Refs.~\cite{Duchesnean:2002,Totsuka:1992}.

\begin{table}[H]
\centering
\caption{\label{tab5}Estimated parameters using Eqs.~(\ref{eq9}) and (\ref{eq10}) with the geometrical combination ($\lambda_1\times {\rm G_1}+\lambda_2\times {\rm G_2}$).} 
\vspace{2mm}
\renewcommand{\arraystretch}{1.0}
\begin{tabular}{cccc}
\hline
  $R_1$ (fm) (G)
& $R_2$ (fm) (G)
& $\lambda_1$
& $\lambda_2$\\
  $0.895\pm 0.017$
& $2.393\pm 0.512$
& $0.498\pm 0.016$
& $0.223\pm 0.075$\\
\hline

  $C$
& $\alpha$ (GeV$^{-1}$) 
& $\beta$ (GeV$^{-1}$) 
& $\chi^2$/ndf ($p$(\%))\\
  $0.938\pm 0.002$
& $2.359\pm 0.056$
& $3.254\pm 0.042$
& 86.3/59 (1.2)\\
\hline
\end{tabular}
\end{table}

\begin{figure}[H]
  \centering
  \vspace{0mm}
  \includegraphics[width=0.55\columnwidth]{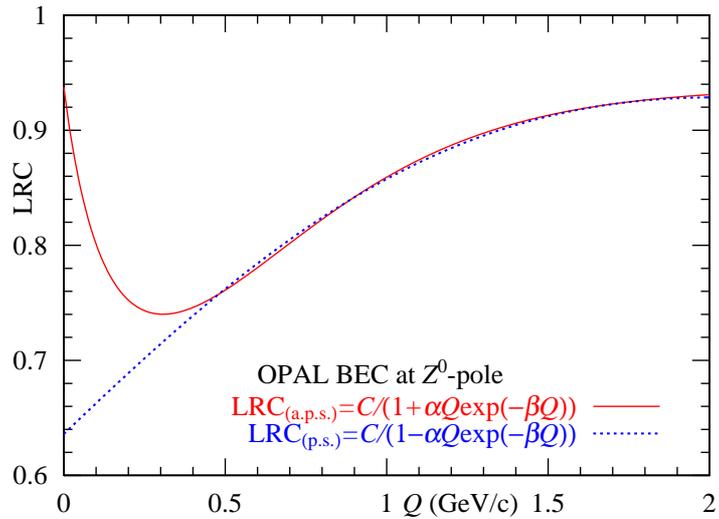}
  \vspace{0mm}
  \caption{\label{fig7}LRCs of the OPAL BEC at the $Z^0$-pole by Eq.~(\ref{eq8}). Numerical values are presented in Tables~\ref{tab4} and \ref{tab5}.}
  \vspace{0mm}
\end{figure}

In Fig.~\ref{fig7}, we compare the two LRC's. The rapidly decreasing behavior of ${\rm LRC_{(a.p.s.)}}$ in the region of $0<Q<0.3$ GeV could be due to the resonance effect of  the $\eta'$ and $\eta$ decay chain, but the coincidence of ${\rm LRC_{(OPAL.)}}$ and  ${\rm LRC_{(a.p.s.)}}$ in the region of $0.5<Q<2.0$ GeV is remarkable.  In this region, ${\rm LRC_{(a.p.s.)}}$ plays the role of ${\rm LRC_{(OPAL.)}}$.

\subsection{\label{sec3-3}Estimation of the contamination due to $\pi$-$K$ pair through $f(Q) = 0.81-0.07Q$ \`a la OPAL}
In conclusion, we corrected for $\pi$-$K$ pair the contamination using the OPAL method, i.e., $f(Q)=0.81-0.07Q$~\cite{Acton:1991xb,Choi:1995}. In Eq.~(\ref{eq7}), $\lambda_1f(Q)$ and $\lambda_2f(Q)$ are used in the analysis. Our results are presented in Table~\ref{tab6}. The $p$-value was improved from 0.98 in Table~\ref{tab5} to 1.34 in Table~\ref{tab6}. Moreover, $\lambda_1+\lambda_2<1.0$ is satisfied in Table~\ref{tab6} in this case. From Table~\ref{tab6}, we obtain the ratio  $R_{\rm 3\mathchar`-all\mathchar`-jet}=17\%$. This figure probably reflects the correlation of the $\pi$-$K$ pair.

\begin{table}[H]
\centering
\caption{\label{tab6}To account contaminating the $\pi$-$K$ pair, we use the OPAL collaboration's function $f(Q)=0.81-0.07Q$. $\lambda_1f(Q)$ and $\lambda_2f(Q)$ are used in the analysis.} 
\vspace{2mm}
\renewcommand{\arraystretch}{1.0}
\begin{tabular}{c|cccc}
\hline
& $R_1$ (fm) (G)
& $R_2$ (fm) (G)
& $\lambda_1$
& $\lambda_2$\\
\lw{${\rm CF_{II}\times LRC_{(a.p.s.)}}$}
& $0.912\pm 0.031$
& $2.574\pm 0.435$
& $0.605\pm 0.026$
& $0.348\pm 0.087$\\
\cline{2-5}

& $C$
& $\alpha$ (GeV$^{-1}$) 
& $\beta$ (GeV$^{-1}$) 
& $\chi^2$/ndf ($p$(\%))\\
& $0.941\pm 0.002$
& $2.222\pm 0.037$
& $3.139\pm 0.046$
& 85.6/59 (1.34)\\
\hline
\hline

& $R$ (fm) 
& 
& $\lambda$ 
& \\
\lw{${\rm CF_{I}\times LRC_{(OPAL)}}$}
& $ 0.944\pm 0.024$
&
& $ 1.091\pm 0.046$
& \\
\cline{2-5}

& $C$
& $\delta$ (GeV$^{-1}$) 
& $\varepsilon$ (GeV$^{-2}$)
& $\chi^2$/ndf ($p$(\%))\\
& $ 0.629\pm 0.010$
& $ 0.490\pm 0.034$
& $-0.126\pm 0.013$
& 113.3/59 (0.0054)\\
\hline
\end{tabular}
\end{table}


\section{\label{sec4}Profile of LRC$_{\rm (a.p.s.)}$ and ($\bm{\lambda}_1$G$_1+\bm{\lambda}_2$G$_2$) in the Euclidean space}
In the previous section, we obtained the concrete values ($\alpha = 2.15\ {\rm GeV^{-1}} = 2.15/5.07$ fm and $\beta = 3.08/5.07$ fm) in ${\rm LRC_{(a.p.s.)}}$. Then, using the following formula~\cite{Shimoda:1993,TSneddon:1995}, we can calculate the profile of (${\rm LRC_{(p.s.)}}-1.0$) in the Euclidean space. Its variable is expressed by $\xi = \sqrt{(x_1-x_2)^2} = \sqrt{(\bm r_1-\bm r_2)^2+(t_1-t_2)^2}$ where $x_1$ and $x_2$ denote the space-time coordinates of two pions. For the behavior of (LRC's$-1$) in the Euclidean space, we have the following density profile of (LRC$-1$) with $s = \pm 1$~\cite{Shimoda:1993,TSneddon:1995,Mizoguchi:2021}
\begin{eqnarray}
\mbox{\Large $\rho$}_{\rm (LRC-1.0)}(\xi) &\!\!\!=&\!\!\! \frac{\xi^2}2 \int_0^{\infty} Q^2 \sum_{k=1}^{\infty} (s)^k \left(\alpha Q e^{-\beta Q}\right)^k J_1(\xi Q)dQ
\nonumber\\
&\!\!\!=&\!\!\! \frac{\xi^2}2 \sum_{k=1}^{\infty} (s\alpha)^k \frac{\Gamma(k+2+2)}{((k\beta)^2+\xi^2)^2)^{(k+2+1)/2}}{\rm P}_{k+2}^{-1}\left(\frac{k\beta}{\sqrt{(k\beta)^2+\xi^2}}\right)
\nonumber\\
&\!\!\!=&\!\!\! \frac{\xi^2}2\left[
(s\alpha)\frac{\Gamma(5)}{(\beta^2+\xi^2)^{2}}{\rm P}_3^{-1}\left(\frac{\beta}{\sqrt{\beta^2+\xi^2}}\right)\right.
\nonumber\\
&\!\!\! &\!\!\! \qquad +(s\alpha)^2\frac{\Gamma(6)}{((2\beta)^2+\xi^2)^{2.5}}{\rm P}_4^{-1}\left(\frac{2\beta}{\sqrt{(2\beta)^2+\xi^2}}\right)
\nonumber\\
&\!\!\! &\!\!\! \qquad \left.
+(s\alpha)^3\frac{\Gamma(7)}{((3\beta)^2+\xi^2)^{3}}{\rm P}_5^{-1}\left(\frac{3\beta}{\sqrt{(3\beta)^2+\xi^2}}\right)
+\cdots
\right],
\label{eq11}
\end{eqnarray}
where ${\rm P}_{k+2}^{-1}$ is the associate Legendre function and ``s'' is the sign function defined as follows,
\begin{eqnarray*}
\left\{
\begin{array}{l}
s=+1\ {\rm for\ LRC_{(p.s)}},\medskip\\
s=-1\ {\rm for\ LRC_{(a.p.s)}}.
\end{array}
\right.\end{eqnarray*}
It is noted that Eq.~({\ref{eq11}) is calculated in the Euclidean space, where the variable\\ $Q_W = \sqrt{(\bm p_1-\bm p_2)^2+(E_1-E_2)^2}$ should be used. The index $W$ means the Wick rotation. However, because we have no such physical quantities described by $Q_W$, we use quantities describe by $Q = \sqrt{(\bm p_1-\bm p_2)^2-(E_1-E_2)^2}$ as an approximate calculation. Thus the variable $\xi$ should be regarded as $D = \sqrt{x^2} = \sqrt{(\bm r_1-\bm r_2)^2-(t_1-t_2)^2}$ in Fig.~\ref{fig8} using a revised Wick rotation, when it is necessary~\cite{Tomonaga:1955,Kozlov:2008}.

Fig.~\ref{fig8} show the density of the profile $\mbox{\Large $\rho$}_{\rm (LRC-1.0)}(\xi)$ because $\int_0^{\infty} Q^2 J_1(\xi Q)dQ$ is divergent. We simultaneously present two Gaussian functions multiplied by the phase space $2\pi^2\xi^3$; 
\begin{eqnarray}
\mbox{\Large $\rho$}_{\rm G}(\xi) =\frac{\xi^3}{8R^4}\exp\left(-\frac{\xi^2}{4R^2}\right).
\label{eq12}
\end{eqnarray}
%

\begin{figure}[H]
  \centering
  \includegraphics[width=0.65\columnwidth]{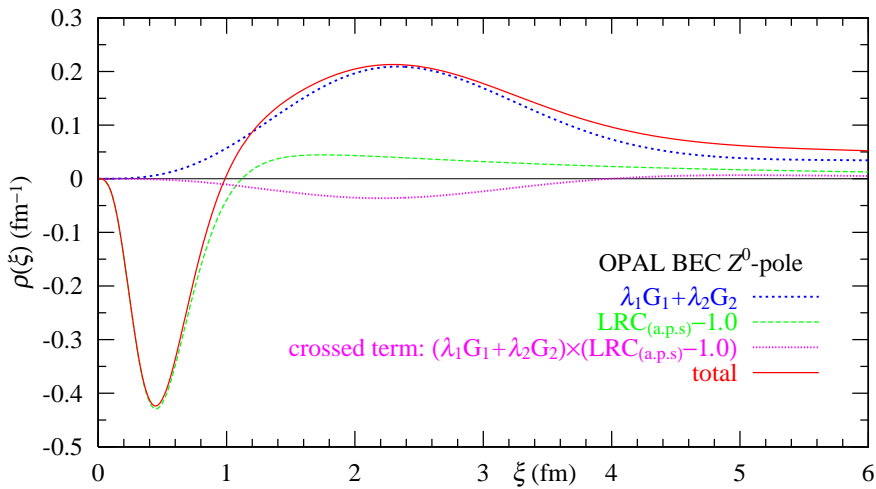}
  \caption{\label{fig8}Pion-pairs density distributions of the OPAL BEC at $Z^0$-pole in four-dimensional Euclidean space, where $\xi$ denotes the distance between two pion-production points. The contribution of the crossed terms ($(\lambda_1{\rm G_1}+\lambda_2{\rm G_2})\times \rho_{\rm (LRC-1)}$), i.e., the probability is small. The value of $C=0.942$ from Table~\ref{tab5} is used.}
\end{figure}

The dip at $\xi = 0.5$ fm is due to the resonance effect of the $\eta'$ and $\eta$ decay chain assigned to LRC mentioned in the subsection \ref{sec3-2}. The role of the dip is the contamination at $Z^0$-pole in $e^+e^-$ collision. The magnitude of the area of the entire profile is $C(\lambda_1+\lambda_2)\cong 1.0$. Note that the contribution of (${\rm LRC_{(p.s.)}}-1.0$) in the entire region ($0< \xi<\infty$) is zero.


\section{\label{sec5}Concluding remarks}
We summarize our present analysis.

\paragraph{C.I:} We propose the second conventional formula with two sources in Eq.~(\ref{eq4}). This formula reflects the physical processes shown in Fig.~\ref{fig2}. There are two types of 2-jet and 3-jet events at $Z^0$-pole. The CF$_{\rm II}$ has a description option. In concrete analysis, the second contribution described by the $\lambda_2\times$exchange function $E_{\rm BE_2}$ is treated as the second approximation. To achieve this goal, we used the inequalities mentioned in Eq.~(\ref{eq5}) in our analysis.

\paragraph{C.II:} We propose an analytic formulation for the LRC with $X = \alpha Q e^{-\beta Q} = \alpha Q(1-\beta Q+\beta^2 Q^2/2!-\cdots)$, with two parameters, $\alpha$ and $\beta$, which are corresponding to parameters $\delta$ and $\varepsilon$ by OPAL collaboration|~\cite{Acton:1991xb}. In our analysis, we use the power series, ${\rm LRC_{(p.s.)}} = \sum_{k=0}^{\infty} X^k$. In the region $0\le Q\le 2$ GeV, ${\rm LRC_{(p.s.)}}$ is almost equivalent with LRC$_{\rm (OPAL)}$ (Fig.~\ref{fig4}).

\paragraph{C.III:} We demonstrated that results with the constraint $\lambda_1 + \lambda_2\le 1.0$ and $\lambda_1 + \lambda_2=$free in Eq.~(\ref{eq4}). In the second analysis including the second term with $\lambda_2$, we assume that the separation of $\lambda_2$ is possible: $\lambda_2=\lambda_2'+\lambda_2''$ and $\lambda_2'=1.0-\lambda_1$. $\lambda_2'$ is the second contribution to BEC. The remaining part $\lambda_2''$ describes the resonance effect due to the $\eta'$ and $\eta$ decay chains. Based on this assumption, we estimate the resonance effect as $\Delta$ in the region $Q\le 0.15\sim 0.3$ GeV in the note column in Table~\ref{tab4}.

\paragraph{C.IV:} We separate the resonance effect due to the $\eta'$ and $\eta$ decay chain in BEC using\\ ${\rm CF_{II}(\lambda_1{\rm G_1}+\lambda_2{\rm G_2})\times LRC_{(p.s.)}}$ analysis. We consider the alternation of ${\rm LRC_{(p.s.)}}$ to improve this separation. To that end, we propose the following formula ${\rm LRC_{(a.p.s.)}}$ in Eq.~(\ref{eq9}). Our result in terms of Eq.~(\ref{eq10}) is shown in Table~\ref{tab5}. The contributions of the 2-jet event and the 3-jet event are estimated in that analysis. The 3-jet to all-jet $R_{\rm 3\mathchar`-all\mathchar`-jet}$ ratio is approximately $0.14=14\%$, which is consistent with the measurement by OPAL collaboration, $R_{\rm 3\mathchar`-all\mathchar`-jet}=17.4\pm 0.7$ \%~\cite{Duchesnean:2002,Totsuka:1992}.

\paragraph{C.V:} Moreover, we followed the method by OPAL collaboration with $f(Q)=0.81-0.07Q$~\cite{Acton:1991xb,Choi:1995}, to subtract the contamination of $\pi$-$K$ pairs,
\begin{eqnarray*}
\lambda f(Q)\times {\rm Gaussian\ function} \longrightarrow \lambda_1 f(Q) {\rm G_1} + \lambda_2 f(Q) {\rm G_2}
\end{eqnarray*}
In the analysis, the sum of the degrees of coherence $\lambda_1$ and $\lambda_2$, $\lambda_1+\lambda_2\approx 0.95 < 1.0$ and the p-value is $p(\%)= 1.34$ is shown in Table~\ref{tab6}. We can compare our result with that of ${\rm CF_{I}\times LRC_{(OPAL)}}$.

\paragraph{C.VI:} Moreover, we investigate the profiles of (${\rm LRC_{(a.p.s.)}} -1.0$) and the exchange functions in Fig.~\ref{fig8}. The profiles of $\left(s\alpha Q e^{-\beta Q}\right)^k$ $(k=1,\, 2,\, 3)$ are calculated in Eq.~(\ref{eq11}). There is an interesting property in $\rho_{(\rm a.p.s)}^{(k)}(\xi)$: The magnitude of the area of $\rho_{(\rm a.p.s)}^{(k)}(\xi)>0$ is the same as that of $\rho_{\rm a.p.s}^{(k)}(\xi)<0$ because of the property of the associate Legendre function. This is remarkable for the physical property of LRCs.

\paragraph{D.I:} The OPAL BEC region is restricted to $0.0<Q<2.0$ GeV. When the upper limit increased, for example, 4.0 GeV, we can obtain more physical information on LRCs (Fig.~\ref{fig7}). The upper value of integration in Eq.~(\ref{eq12}) is infinity. If we replace it with $Q=4$ GeV, we obtain almost the same figures in Fig.~\ref{fig8}.

\paragraph{D.II:} We would like to analyze the other BECs at 0.9 TeV and 7 TeV with CMS collaboration using Eq.~(\ref{eq4}) in the future.\\

\noindent
{\it Acknowledgments.} One of authors (M.B.) would like to thank his colleagues at Department of Physics in Shinshu University.


\end{document}